\documentclass{svjour3}                     
\smartqed  
\usepackage{graphicx}
\journalname{jltp}
\begin{document}

\title{Competition between electronic cooling and Andreev dissipation in a superconducting micro-cooler}

\author{Sukumar Rajauria \and P. Gandit \and F.W.J. Hekking \and B. Pannetier \and H. Courtois}

\institute{Sukumar Rajauria \and P. Gandit  \and B. Pannetier \at
Institut N\' eel, CNRS and Universit\'e Joseph Fourier, 25 Avenue des Martyrs, BP 166, 38042 Grenoble, France.
\and F. W. J. Hekking \at
LPMMC, Universit\'e Joseph Fourier and CNRS, 25 Avenue des Martyrs, BP 166, 38042 Grenoble, France.
\and H. Courtois \at 
Low Temperature Laboratory, Helsinki University of Technology, P.O. Box 3500, 02015 TKK, Finland; Institut N\' eel, CNRS and Universit\'e Joseph Fourier, 25 Avenue des Martyrs, BP 166, 38042 Grenoble, France. \email{herve.courtois@grenoble.cnrs.fr}}
\date{Received: date / Accepted: date}

\maketitle

\begin{abstract}
We discuss very low temperature experiments on superconducting micro-coolers made of a double Normal metal - Insulator - Superconductor junction. We investigate with a high resolution the differential conductance of the micro-cooler as well as of additional probe junctions. There is an explicit crossover between the single quasi-particle current and the phase-coherent Andreev current. We establish a thermal model by considering the thermal contribution due to the Andreev current. The related increase of the electron temperature is discussed, including the influence of several parameters like the phase-coherence length or the tunnel junction transparency.
\end{abstract}

\section{Introduction}

The transfer of quasi-particles across the junction between a Normal metal (N) and a Superconductor (S) is mainly governed by two processes.

Single quasi-particles can tunnel from the normal metal to the superconductor if their energy $E$ compared to the superconductor Fermi level is larger than the superconducting gap $\Delta$ ($E > \Delta$). This energy selectivity induces a cooling of the electronic population of the normal metal \cite{RMP-Giazotto} in a S-I-N (where I stands for Insulator) junction biased at a voltage below the gap $\Delta/e$. As the heat current direction does not depend on the sign of the bias, S-I-N-I-S micro-coolers based on a double tunnel junction feature a double cooling power and an improved efficiency due to the better thermal isolation of the metal. The electronic temperature reduction reaches an optimum at a voltage bias just below the gap. In a Al-based device, normal metal electrons can cool from a bath temperature of 300~mK down to below 100~mK \cite{RMP-Giazotto,APL-Nahum}.

For an energy $E$ below the gap ($E < \Delta$), two quasi-particles can tunnel into the superconductor and form a Cooper pair in the superconductor. This mechanism is called Andreev reflection since it can be viewed as the reflection of an electron into a hole \cite{Andreev}. The Andreev current is widely believed to be a dissipation-less current, which means that it would contribute only as a charge current.

The junction normal-state resistance $R_N$ is proportional to $1/T$, where $T$ is the single quasi-particule tunneling probability. In comparison, the Andreev reflection is a two-particule process and its probability is proportional to $T^2$. In a ballistic picture \cite{BTK}, Andreev reflection is therefore vanishing in a S-I-N tunnel junction, where the interface transparency is small.

In the presence of disorder, the electrons specularly reflected at the superconducting interface are confined in the vicinity of the barrier and hit the interface several times. If phase coherence is preserved, the probability amplitude of Andreev reflection for every attempt add constructively \cite{PRL-vanWees}. As a hole and an electron travel on the same trajectory but in opposite directions, this addition is immune to the phase randomization induced by the disorder. At an energy $E$ compared to the Fermi level in the superconductor, the constructive coherent addition of probability builds up over the energy-dependent coherence length:
\begin{equation}
L_{E}=\sqrt{\frac{\hbar D}{E}}
\end{equation}
or by the phase-coherence length $L_{\varphi}$ if smaller \cite{Superlattices-Courtois}. If one considers a electronic population at thermal equilibrium and a small bias $eV \ll k_BT$, one can define a mean coherence length:
\begin{equation}
L_T=\sqrt{\frac{\hbar D}{2 \pi k_B T}}.
\end{equation}

With the effect of phase-coherent confinement taken into account, the Andreev channel contributes as a conductance of the order of $R_{diff}/R_N^2$, where $R_{diff}$ is the resistance of the diffusive phase-coherent normal metal. At low enough temperature and bias, it can dominate the conductance of S-I-N junctions with an intermediate transparency. A zero-bias conductance peak is then observed, with a width given by the Thouless energy of the coherent diffusive region, or the thermal energy $2k_BT$ if larger. The enhanced Andreev current \cite{Physica-Volkov,PRL-Hekking} due the phase-coherent confinement by the disorder was first observed in Ref. \cite{PRL-Kastalsky}. It was later shown that the Andreev current can be modulated by a magnetic flux \cite{PRL-Pothier}. In the limit of a strong confinement, for example due to a second barrier within the normal metal \cite{PRB-Quirion}, this effect is more often called reflectionless tunneling \cite{PRL-Beenakker} as the specular reflection at a tunnel barrier appears as reduced.

\section{Experimental results}

\begin{figure}[t]
\centering
\includegraphics[width=0.68\linewidth]{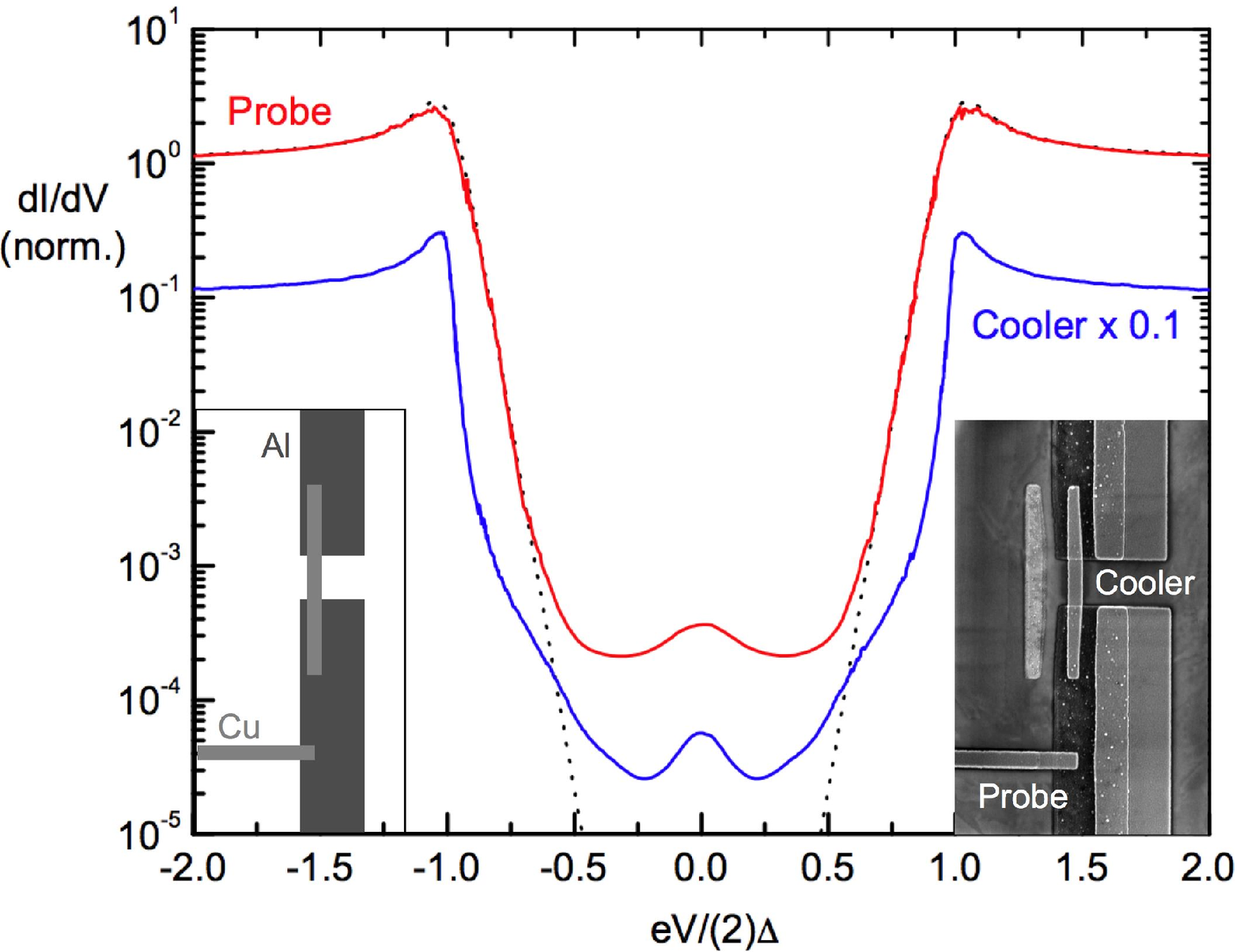}
\includegraphics[width=0.31\linewidth]{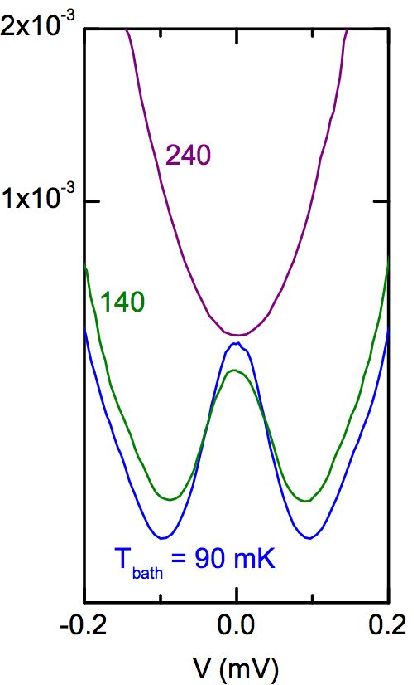}
\caption{(Color online) Left: Differential conductances measured at a cryostat temperature of 90 mK. Top (red) curve: data of one probe junction 1.55 $\mu$m from the cooler junction and of normal-state resistance $R_N$ = 2.76 k$\Omega$. The black dotted line is a fit to Eq. \ref{eq:symmetric-NIS} describing the single quasi-particle tunneling current. Bottom (blue) curve: cooler junction data with a normal state resistance $R_{N}$ = 1.9 k$\Omega$. Inset: Schematics and micrograph of a cooler made of two Al-AlO(x)-Cu junctions in series. The area of a cooler junction is 1.5 $\times$ 0.3 $\mu$m$^2$. In addition to the cooler, one of the three Al-AlO$_x$-Cu probe junctions on the bottom superconducting electrode is visible. The superconducting gap is $2 \Delta$ = 0.43 meV. The voltage axis is normalized to $\Delta$ (probe data) or $2 \Delta$ (cooler data). Right: Normalized differential conductance of the cooler junction as a function of voltage at different cryostat temperatures: 240 (purple), 140 (green) and 90 mK (blue line).}
\label{fig:90mK-cooler-probe}
\end{figure}

Fig. \ref{fig:90mK-cooler-probe} left (inset) shows the micrograph of a typical cooler device, where a central normal metal Cu island is attached to two superconducting Al reservoirs through tunnel junctions. The two 40 nm thick and 1.5 $\mu$m wide superconducting Al electrodes were in-situ oxidized in 0.2 mbar of oxygen for 3 min before the deposition of the central Cu island, which is 4 $\mu m$ long, 0.3 $\mu$m wide and 50 nm thick. In addition to these cooler junctions, we added three Cu tunnel probes of area 0.3 $\times$ 0.3 $\mu$m$^2$ on one Al electrode. Due to the large volume of the probe Cu electrode, the probe is strongly thermalized to the cryostat temperature. In the following, we will describe the experimental results obtained on one sample, while we observed a very similar behavior in three more samples. At intermediate temperature (above about 200 mK), these samples showed a behavior identical to the one reported in Ref. \cite{PRL-Sukumar1}. The charge current in both junctions can then be described by the sole single quasi-particles contribution. The cooler shows an electronic cooling, while the probe does not, as expected. 

Let us now concentrate on the very low temperature regime. Fig. \ref{fig:90mK-cooler-probe} left shows the differential conductance of the cooler double junction and of one of the probe junction at the cryostat temperature of 90 mK. It was obtained by numerical differenciation of the measured current-voltage characteristics. At intermediate bias, the probe junction data is well fitted (black dotted line) by considering only the single quasi-particle tunnel current given by:
\begin{equation}
I_T(V) = \frac{1}{eR_{N}}\int^{\infty}_{0} N_{S}(E)[f_{N}(E - eV) -f_{N}(E + eV)] dE,
\label{eq:symmetric-NIS}
\end{equation}
\cite{ULTI} with an electronic temperature of 105 mK, which is slightly higher than the cryostat temperature.

Close to zero bias, the cooler and the probe feature a peak in the differential conductance. This effect cannot be accounted for by a linear leakage as it would lead to a saturation of the conductance near zero bias. As discussed in Ref. \cite{ULTI}, this zero bias anomaly cannot be fitted by considering a non-equilibrium distribution in the normal metal or by considering a smeared density of states in the superconducting electrodes \cite{PRL-Pekola}. Below about 200 mK, the zero bias conductance increases when the cryostat temperature is lowered (see Fig. \ref{fig:90mK-cooler-probe} right), which suggests that it is a phase-coherent effect. In the cooler junction, the differential conductance is decreased at intermediate bias and more peaked at the gap edge compared to the one of the probe junction, which exemplifies the cooling of the normal metal island electrons.

\section{The Andreev current at thermal equilibrium}

In the following, we will ascribe the zero bias enhancement of the differential conductance to a phase-coherent Andreev current.  We will use the theory of Ref. \cite{PRL-Hekking,PRB-Hekking}. In our samples, the coherence length $L_{T}$ at $T$ = 90 mK is about 0.33 $\mu$m, which is of the order of the junction dimensions. At low energy, the propagation of the Cooperon in the electrode is cut by the phase-coherence length $L_{\varphi}$, which is expected to be about 2 $\mu$m, \mbox{i.e.} larger than the junction dimensions. Therefore, we will use the 1D regime for the diffusion of the Cooperon in both the normal metal and the superconductor electrode. We take into account the finite gap of superconductor, so that the calculation is valid for $eV,kT<\Delta$, and we include the disorder both in the normal metal and in the superconductor \cite{Vasenko}. 

\begin{figure}[t]
\centering
\includegraphics[width=0.5\linewidth]{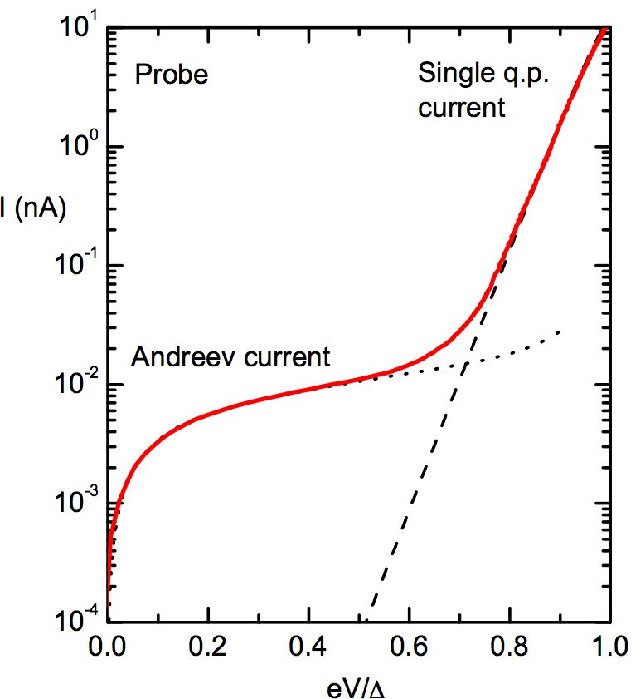}
\includegraphics[width=0.492\linewidth]{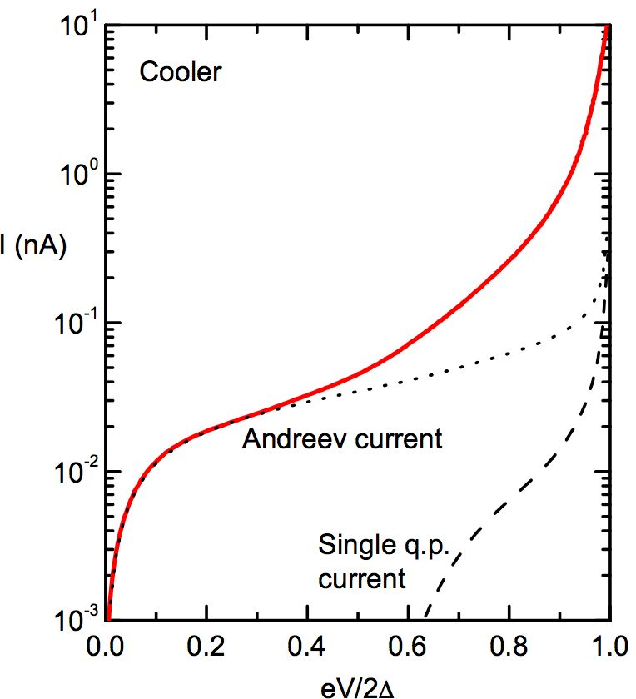}
\caption{(Color online) Experimental current-voltage characteristics (full red lines) of the probe (left) and cooler (right) junctions at a cryostat temperature of 90 mK, compared to calculated curves. The dotted lines are fits with the calculated phase-coherent Andreev current and the dashed lines shows the fit with the calculated single quasi-particle current. The parameters are: D = 80~cm$^{2}$/s, $L_{\varphi}$ = 1.5 $\mu$m, $M$ = 1.37 (probe) or 0.49 (cooler), $2 \Delta$ = 0.43 meV, $K A$ = 144 W.K$^{-4}$. }
\label{fig:exptprobe-Andreev-quasi}
\label{fig:90mK-folklore}
\end{figure}

We have fitted the probe data by calculating the sum of the single quasi-particle current $I_{T}$ and the phase-coherent Andreev current $I_{A}$. Fig. \ref{fig:exptprobe-Andreev-quasi} left shows the excellent fit with the two contributions of the Andreev current at low bias and of the single quasi-particle current at higher bias. The fit parameters are: $L_{\varphi}$ = 1.5 $\mu$m, $\Delta$ = 0.228 meV, T = 105 mK. We took the measured values of the diffusion coefficient D = 80 cm$^{2}$/s and $R_{N}$ = 2.76 k$\Omega$. We have used the same parameters to fit successfully the experimental data from the two other probe junctions. In the fit, we had to scale the phase-coherent Andreev current by a multiplying factor $M$ = 1.37. This factor \cite{PRL-Pothier} could be due to small inhomogeneities in the tunnel barrier, which are not considered here. 

\section{The behavior of the cooler junction in the presence of an Andreev current}

In order to understand the behavior of the cooler junction, we need to consider the heat balance in the central normal metal.  Here we assume a quasi-equilibrium situation: the electrons and the phonons in the metallic island follow a thermal distribution function at a respective temperature $T_{e}$ and $T_{ph}$, which are in general different from the bath temperature $T_{bath}$ of the cryostat. 

The single quasi-particle current is responsible for the cooling power:
\begin{equation}
P_{cool}=\frac{1}{e^2R_{n}}\int^{\infty}_{-\infty}(E-\frac{eV}{2})n_{s}(E)[f_{N}(E-\frac{eV}{2})-f_{S}(E)]dE
\end{equation}
The cooling power is compensated by the electron-phonon coupling power 
\begin{equation}
P_{e-ph}=\Sigma U(T_{e}^{5} - T_{ph}^{5}),
\end{equation}
so that $2P_{cool} + P_{e-ph} = 0$, where the factor 2 is due to the fact that we have N-I-S junction in series. We consider that the heat given to the normal metal phonons is compensated by the Kapitza coupling with the phonons of the substrate kept at the bath temperature $T_{bath}$: $P_{K}= KA(T_{bath}^{4}-T_{ph}^{4})$, so that $P_{e-ph} + P_{K} = 0$. Here $P_K$ is the power flow through the Kapitza resistance, $K$ is the Kapitza parameter depending on the materials in contact, and $A$ is the contact area. The Kapitza thermal resistance is significant for intermediate temperatures and above ($T >$ 300 mK), which can lead to the cooling of the normal metal phonons \cite{PRL-Sukumar1}. We have assumed that the Andreev current does not dissipate any heat. 

To fit the experiment, we first solve numerically the thermal model discussed above for a relatively high temperature ($T >$ 300 mK), where the contribution of the phase-coherent Andreev current is negligible. We take the electron-phonon coupling coefficient $\Sigma$ = 2 nW.$\mu$m$^{-3}$.K$^{-5}$ and obtain from the fit the Kapitza coupling parameter $K A$ = 144 W.K$^{-4}$. The Kapitza coefficient $K$ is comparable to the one found in our previous experiments \cite{PRL-Sukumar1}.

We turn afterwards to the very low temperature regime of interest here. Fig. \ref{fig:90mK-folklore} right shows the direct current-voltage characteristic obtained from the cooler junction (full red line) along with calculated curves (dashed and dotted lines). The dashed line is the calculated current-voltage characteristic at a 90 mK cryostat temperature including the charge and heat currents of the single quasi-particle tunneling only. The agreement is poor, which confirms the need to include the Andreev current contribution to the junction. The dotted line shows the result of the calculation based on the thermal model with the charge current given by the sum of the single quasi-particle current and the Andreev current. The fit parameters are the same than for the probe except for the scaling factor $M$ = 0.49. The difference with the probe junction factor is not understood, although it could be due to the difference in geometry between the two junctions. The addition of the phase-coherent current provides an acceptable fit at low bias but shows a clear discrepancy at intermediate voltage. The experimental curve predicts a larger current than what is obtained from the thermal model. This demonstrates that an excess dissipation term or an extra current contribution is missing in the thermal model.

\begin{figure}[t]
\centering
\includegraphics[width=0.7\linewidth]{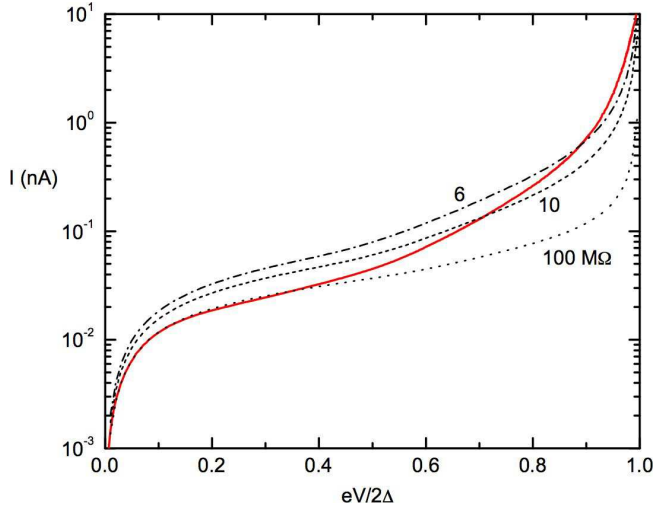}
\caption{(Color online) Current-voltage characteristic of the cooler junction (full red line) along with curves calculated with the calculated thermal model and including an excess leakage due to a linear resistance of 6 (dash-dotted line), 10 (dashed line) and 100 M$\Omega$ (dotted line).}
\label{fig:90mK-cooler-leakage}
\end{figure}

As a possible explanation, Fig. \ref{fig:90mK-cooler-leakage} shows the comparison between the experiment and a calculated curve from the thermal model, which includes an additional dissipation due to a linear resistance. Here the leakage contributes both as a dissipation in the normal metal and as a current across the junction. The differential conductance at zero bias of the cooler junction (see Fig. \ref{fig:90mK-cooler-probe}) gives a minimum leakage resistance of 20 M$\Omega$. Fig. \ref{fig:90mK-cooler-leakage} shows that adding such an extra dissipation term to the thermal model discussed above does not provide a good description of the experiment. 

\section{The Andreev current induced dissipation}

Let us now discuss the heat transfer due to the Andreev current. The work performed by the current source feeding the circuit with the extra Andreev current $I_{A}$ generates a Joule heat $P_{A} = I_{A} V$ that is deposited in the normal metal \cite{PRB-Averin}. This heat is deposited entirely in the normal metal and does not perturb the superconductor. Hence, the net cooling power of the S-I-N junction is reduced and can be re-defined as $P_{cool} - P_{A}$.

\begin{figure}[t]
\centering
\includegraphics[width=0.7\linewidth]{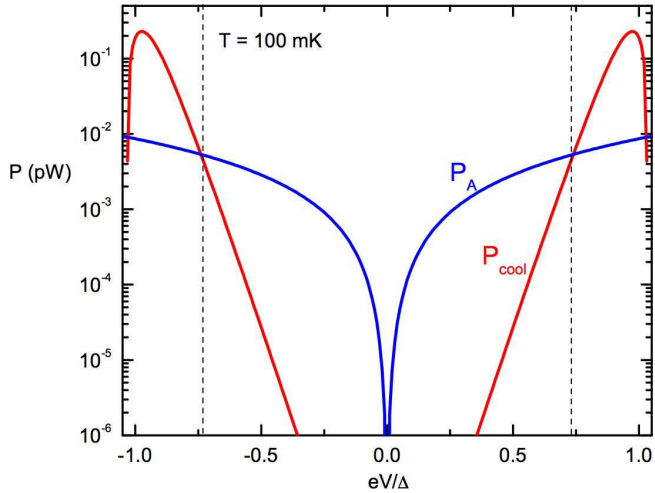}
\caption{(Color online) Calculated power due to quasi-particle cooling P$_{cool}$ (red line) and the heat dissipation due to the Andreev current $P_{A}$ (blue line) as a function of voltage bias for a N-I-S junction at $T_{bath}$ = 100 mK. The parameters used in the calculation are the same as in Fig. \ref{fig:90mK-folklore}.}
\label{fig:power-compare}
\end{figure}

Fig. \ref{fig:power-compare} shows the quantitative comparison at 100 mK of the cooling power $P_{cool}$ due to single quasi-particle tunneling (red line) and the dissipation $P_{A}$ due to the Andreev current (blue line). Near zero bias, $P_{cool}$ is almost zero due to the absence of quasi-particles. It attains its maximum near the gap. The Andreev current induced dissipation $P_A$  increases sharply near zero bias. Close to the gap voltage, the cooling power out-does the Andreev dissipation. As the latter depends strongly on the transparency of the junction, it surpasses the single quasi-particle cooling at a varying temperature and bias. In the present device with a tunnel barrier transparency of about 10$^{-5}$, the Andreev current dissipation becomes relevant only below 200 mK.

\begin{figure}[t]
\centering
\includegraphics[width=0.7\linewidth]{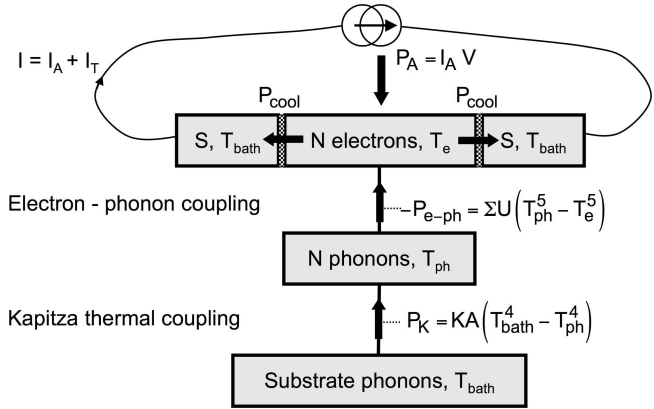}
\caption{Schematic of the complete thermal model of S-I-N-I-S micro-coolers at very low temperature (see text).}
\label{fig:model}
\end{figure}

\begin{figure}[t]
\centering
\includegraphics[width=0.67\linewidth]{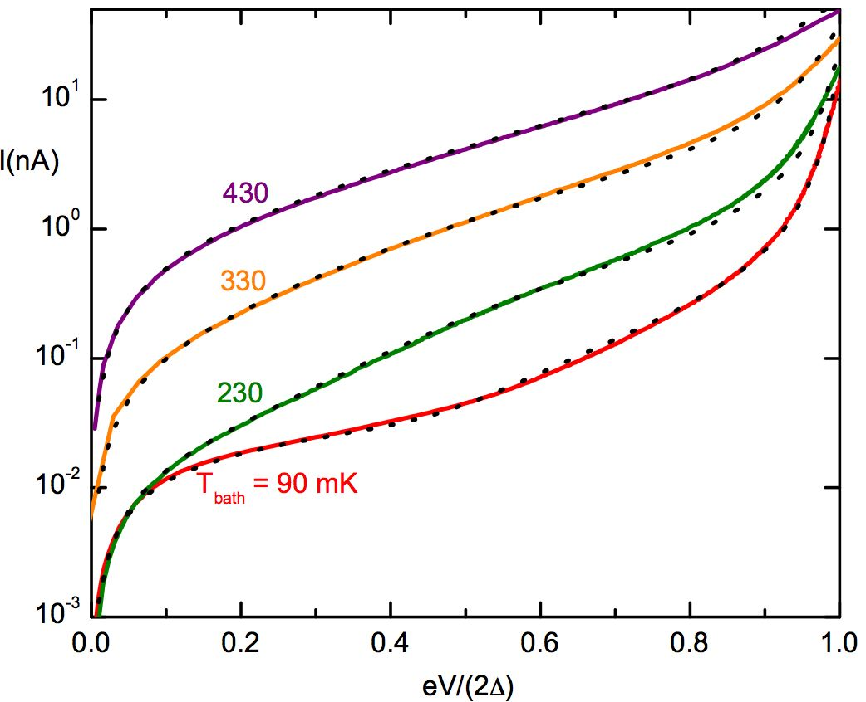}
\includegraphics[width=0.32\linewidth]{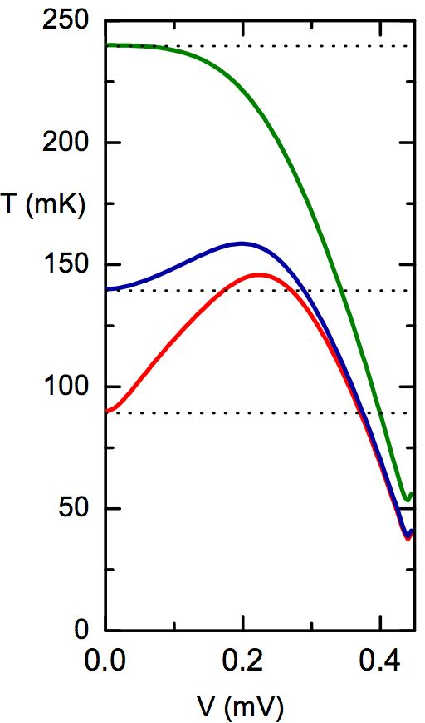}
\caption{(Color online) Left: Current voltage characteristic of the cooler junction at different cryostat temperatures $T_{bath}$ together with the calculated best fit from the full thermal model including the charge and heat contribution due to the Andreev current. The temperatures are from top to bottom 430 (purple), 330 (orange), 230 (green) and 90 mK (red line). Right: Dependence of the calculated electronic temperature with the voltage with the parameters obtained from the fit to the experiment and for a series of cryostat temperatures: 230 (green line), 140 (blue) and 90 mK (red).}
\label{fig:I-V fit}
\label{fig:T-V}
\end{figure}

Fig. \ref{fig:model} shows the schematic of the full thermal model of the device. Here we have included the work $I_A V$ done on the central metallic island by the current source. At steady state, the heat balance for the normal metal electrons can be rewritten as:
\begin{equation}
2P_{cool} + P_{e-ph} - I_{A}V = 0.
\label{Eq:complete-electron}
\end{equation}
With this complete heat balance equation taken into account, we solve the thermal model and calculate the total contribution to the current in the cooler junction. Fig. \ref{fig:I-V fit} left shows the comparison of the experiment (full colored lines) and the thermal model (dotted lines). The agreement is very good at every accessible cryostat temperature. The Andreev current contribution becomes negligible compared to the single quasi-particle current contribution at temperatures above 200 mK.

The above conclusion on the Andreev heat is independent of the phonon cooling \cite{PRL-Sukumar1}. Assuming the perfect thermalization of the phonons to the substrate temperature would change the total calculated current by less than 2 $\%$ at 90 mK, which means that phonon cooling has a negligible role in the data analysis at very low temperature. This is consistent with the expected negligible amplitude of the phonon cooling in this temperature range.

The above fit with the thermal model also provides us with the electron temperature for every bias. Fig. \ref{fig:T-V} right shows the calculated electron temperature in the central metallic island as a function of voltage bias across the cooler junction and for different cryostat temperatures $T_{bath}$ = 90, 140 and 230 mK. At very low temperature, the electron temperature first increases with the bias as the Andreev current-induced dissipation is dominant. When the voltage bias approaches the gap, the single quasi-particle tunneling based cooling dominates. As the bath temperature increases, the Andreev current induced dissipation becomes less effective and for $T_{bath}$ = 230 mK the electronic cooling always prevails \cite{PRL-Sukumar2}.

Although the Andreev reflection is a small effect as a charge current in S-I-N-I-S cooling devices, the related heat contribution is extremely efficient at very low temperature. A basic explanation for this is the following. At very low temperature the quasi-particle based cooling power has a small efficiency compared to the Joule power $IV$. It is of the order of $T_{e}/\Delta$, which is about 5$\%$ at a 100 mK electron temperature. In contrast, the Andreev current induced dissipation is the full Joule power.

\section{Conclusion}

As a summary, we observed a peak in the differential conductance at low bias in the probe and cooler junctions of superconducting micro-coolers, which is due to a phase-coherent Andreev current. A quantitative thermal analysis of the micro-cooler behavior demonstrates the importance of the dissipation induced by the Andreev current, as it dominates the cooling power at very low temperature over a significant voltage range.

The above conclusion poses a challenge to diminish the dissipation induced by the phase-coherent Andreev current in S-I-N-I-S micro-cooling devices. What are the right parameters needed to optimize the cooling power? The induced dissipation $P_A$ due to the Andreev current depends strongly on the transparency. It scales as $1/R_{N}^2$ whereas the quasi-particle cooling $P_{cool}$ scales as $1/R_{N}$. A resistance optimum in terms of cooling power can then be found for every temperature. As the temperature decreases, the high-energy tail of the electron energy distribution is vanishing, which reduces the cooling. In contrast, the Andreev current would increase. It is thus expected that the optimum resistance increases when the electronic temperature decreases. This qualitative discussion needs to be completed by a more quantitative calculation.

For a fixed tunnel resistance, a smaller phase-coherence length  $L_{\varphi}$ would diminish the phase-coherent Andreev current as well as the related dissipation. For instance, a disordered material would be a better choice as a normal metal since it would have a much shorter phase-coherence length than a pure metal like Copper. The latter strategy seems clearly promising for new devices with an improved efficiency at very low temperatures.

\section{Acknowledgements}

We acknowledge the financial support from the ANR contract "Elec-EPR", the ULTI-3 and the NanoSciERA "Nanofridge" EU projects. H. Courtois thanks the Low Temperature Laboratory for hospitality. The authors thank A. Vasenko and M. Houzet for discussions.


\begin{thebibliography}{}
\bibitem{RMP-Giazotto} F. Giazotto, T. T. Heikkil\"a, A. Luukanen, A. M. Savin and J. P. Pekola,  Rev. Mod. Phys. {\bf 78}, 217 (2006).
\bibitem{APL-Nahum} M. Nahum, T. M. Eiles and J. M. Martinis, Appl. Phys. Lett. {\bf 65}, 3124 (1994).
\bibitem{Andreev} A. F. Andreev, Zh. Eksp. Teor. Fiz. {\bf 46}, 1823 (1964); D. Saint-James, J. Phys. France {\bf 25}, 899 (1964).
\bibitem{BTK} G. E. Blonder, M. Tinkham, and T. M. Klapwijk, Phys. Rev. B {\bf  25}, 4515 (1982).
\bibitem{PRL-vanWees} B. J. van Wees, P. de Vries, P. Magn\'ee, and T. M. Klapwijk, Phys. Rev. Lett. {\bf 69}, 510 (1992).
\bibitem{Superlattices-Courtois} H. Courtois, P. Gandit, B. Pannetier, and D. Mailly, Superlatt. and Microstruct. {\bf 25}, 721 (1999).
\bibitem{Physica-Volkov} A. F. Volkov, A. V. Zaitsev, and T.M. Klapwijk, Physica C {\bf 210}, 21 (1993).
\bibitem{PRL-Hekking} F. W. J. Hekking and Yu. V. Nazarov, Phys. Rev. Lett. {\bf 71}, 1625 (1993).
\bibitem{PRL-Kastalsky} A. Kastalsky, A. W. Kleinsasser, L. H. Greene, R. Bhat, F. P. Milliken, and J. P. Harbison, Phys. Rev. Lett. {\bf  67}, 3026 (1991).
\bibitem{PRL-Pothier} H. Pothier, S. Gu\'eron, D. Est\`eve, and M. H. Devoret, Phys. Rev. Lett. {\bf 73}, 2488 (1994).
\bibitem{PRB-Quirion} D. Quirion, C. Hoffmann, F. Lefloch and M. Sanquer, Phys. Rev. B {\bf 65}, 100508 (2002).
\bibitem{PRL-Beenakker} C. W. Beenakker, B. Rejaei, and J. A. Melsen, Phys. Rev. Lett. {\bf 72}, 2470 (1994).
\bibitem{PRL-Sukumar1} Sukumar Rajauria, P. S. Luo, T. Fournier, F. W. J. Hekking, H. Courtois, and B. Pannetier, Phys. Rev. Lett. {\bf 99}, 047004 (2007).
\bibitem{ULTI} H. Courtois, Sukumar Rajauria, P. Gandit, F. W. J. Hekking, and B. Pannetier, J. of Low Temp. Phys. {\bf 153}, 325 (2008).
\bibitem{PRL-Pekola} J. P. Pekola, T. T. Heikkil\"a, A. M. Savin, J. T. Flyktman, F. Giazotto, and F. W. J. Hekking, Phys. Rev. Lett. {\bf 92}, 056804 (2004).
\bibitem{PRB-Hekking} F. W. J. Hekking and Yu. V. Nazarov, Phys. Rev. B {\bf 49}, 6847 (1994).
\bibitem{Vasenko} A. Vasenko, S. Rajauria, H. Courtois and F. W. J. Hekking, unpublished (2008).
\bibitem{PRB-Averin} A. Bardas and D. Averin, Phys. Rev. B {\bf 52}, 12873 (1995).
\bibitem{PRL-Sukumar2} Sukumar Rajauria, P. Gandit, T. Fournier, F. W. J. Hekking, B. Pannetier, and H. Courtois, Phys. Rev. Lett. {\bf 100}, 207002 (2008).
\end{thebibliography}
\end{document}